\documentclass[aip,reprint,superscriptaddress,nofootinbib,twocolumn,10pt]{revtex4-2}

\usepackage[linktoc=page,colorlinks,urlcolor=blue,citecolor=blue,linkcolor=blue]{hyperref}
\usepackage{amssymb,amsmath}
\usepackage{upgreek}
\usepackage{graphicx}
\usepackage{natbib}
\usepackage{lineno}
\usepackage[dvipsnames]{xcolor}
\usepackage{overpic}
\usepackage[font=small,labelfont=bf,
   justification=raggedright,
   format=plain]{caption}
\newcommand{\snl}{Sandia National Laboratories, Albuquerque, New Mexico 87185, USA}
\newcommand{\cint}{Center for Integrated Nanotechnologies, Sandia National Laboratories, Albuquerque, New Mexico 87123, USA}
\newcommand{\unm}{University of New Mexico Department of Physics and Astronomy, Albuquerque, New Mexico 87131, USA}

\usepackage{verbatim}

\begin{document}
\title{Current Paths in an Atomic Precision Advanced Manufactured Device Imaged by Nitrogen-Vacancy Diamond Magnetic Microscopy}

\date{\today}
\author{Luca Basso}\affiliation{\cint}
\author{Pauli Kehayias}\affiliation{\snl}
\author{Jacob Henshaw}\affiliation{\cint}
\author{Maziar Saleh Ziabari}\affiliation{\cint}\affiliation{\unm}
\author{Heejun Byeon}\affiliation{\cint}
\author{Michael P. Lilly}\affiliation{\cint}
\author{Ezra Bussmann}\affiliation{\snl}
\author{Deanna M. Campbell}\affiliation{\snl}
\author{Shashank Misra}\affiliation{\snl}
\author{Andrew M. Mounce}\affiliation{\cint}

\begin{abstract}
The recently-developed ability to control phosphorous-doping of silicon at an atomic level using scanning tunneling microscopy (STM), a technique known as atomic-precision-advanced-manufacturing (APAM), has allowed us to tailor electronic devices with atomic precision, and thus has emerged as a way to explore new possibilities in Si electronics. In these applications, critical questions include where current flow is actually occurring in or near APAM structures as well as whether leakage currents are present. In general, detection and mapping of current flow in APAM structures are valuable diagnostic tools to obtain reliable devices in digital-enhanced applications. In this paper, we performed nitrogen-vacancy (NV) wide-field magnetic imaging of stray magnetic fields from surface current densities flowing in an APAM test device over a mm-field of view with $\upmu$m-resolution. To do this, we integrated a diamond having a surface NV ensemble with the device (patterned in two parallel mm-sized ribbons), then mapped the magnetic field from the DC current injected in the APAM device in a home-built NV wide-field microscope. The 2D magnetic field maps were used to reconstruct the surface current density, allowing us to obtain information on current paths, device failures such as choke points where current flow is impeded, and current leakages outside the APAM-defined P-doped regions. Analysis on the current density reconstructed map showed a projected sensitivity of $\sim$0.03 A/m, corresponding to a smallest detectable current in the 200 $\upmu$m-wide APAM ribbon of $\sim$6 $\upmu$A. These results demonstrate the failure analysis capability of NV wide-field magnetometry for APAM materials, opening the possibility to investigate other cutting-edge microelectronic devices.
\end{abstract}

\maketitle
\section{Introduction}
Atomic-precision control over crystalline Si synthesis and processing are putting pure quantum mechanics phenomena (e.g.~spin and tunnel effects) on the Si device engineer’s palette of reliable options for creating new microelectronics. One recent addition~\cite{ruess2004, shen2004} to the palette is a technique known as atomic-precision advanced manufacturing (APAM). A typical APAM process first creates a lithographic pattern of Si dangling bonds on a hydrogen-passivated crystalline Si surface, using scanning tunneling microscopy (STM). Next, the Si surface is exposed to phosphine gas that selectively adsorbs on sites where Si dangling bonds have been exposed, leading to atomically-precise planar structures made of P-donors~\cite{simmons2005, ruess2007}. APAM allows us to dope Si in full 2D atomic-precision~\cite{mckibbin2013,ward2017,skeren2018,ward2020,bussmann2021}, to form structures such as wires~\cite{weber2012,mckibbin2013}, quantum dots~\cite{fuhrer2009}, and few- and single-dopant features~\cite{fuechsle2012,he2019} with lattice-site (3.8 \AA) control~\cite{bussmann2021}. Doped features can be positioned, arrayed, coupled, and gated with atomic precision to form tunnel junctions~\cite{wang2020}, single-atom transistors~\cite{fuechsle2012}, and a host of other possibilities, all compatible with near-standard CMOS manufacturing~\cite{ward2017,skeren2018,skeren2020}. This gives us an exciting tool to explore far-reaching ideas, such as quantum computing~\cite{ward2020}, quantum simulation~\cite{wang2021}, and digital electronics that incorporate controlled quantum effects into Si foundry-compatible platforms~\cite{ward2017,ward2020,skeren2018,gao2020,lu2021}. 

Currently, APAM devices work only in cryogenic conditions, hence their use is limited to applications where low temperatures are required (e.g.~in quantum computing) to achieve long coherence times. Realizing broader possible applications for APAM techniques in room-temperature conventional electronics is a big challenge, as it requires understanding of carrier transport in ultrathin high-density P-doped Si structures. Electronic transport in APAM-doped materials and structures has been investigated and understood at cryogenic conditions ($T <$ 4.2 K)~\cite{Mamaluy2021}. APAM-doped regions would dominate carrier transport just by virtue of the high APAM doping levels ($10^{22}$ cm$^{-3}$), while doping in surrounding Si was kept sufficiently low ($\ll 3\cdot 10^{18}$ cm$^{-3}$, i.e.~the metal-insulator transition) that leakage paths through the substrate and cap layers are frozen out at 4 K, contributing little to a device's transport properties. In higher-T applications, where substrate freeze-out does not occur, unintended leakage currents can be much more than a nuisance background power-drain, and can potentially swamp the intended current channels and mask the intended device functionality~\cite{gao2020}. Thus, understanding and engineering charge carrier confinement to intended doped transport pathways is an integral challenge particularly relevant for room-temperature quantum-enhanced digital electronics. As a result, methods to detect and map the carrier flow in APAM-doped structures are valuable tools to build the understanding necessary to achieve reliable devices. More generally, a broader suite of failure analysis tools validated not only for APAM, but for a wide range of novel materials, is timely and relevant~\cite{bussmann2015,scrymgeour2017,gramse2017, katzenmeyer2020, gramse2020, kolker2022}. 

\begin{figure*}[ht]
\centering
\includegraphics[width=0.9\linewidth]{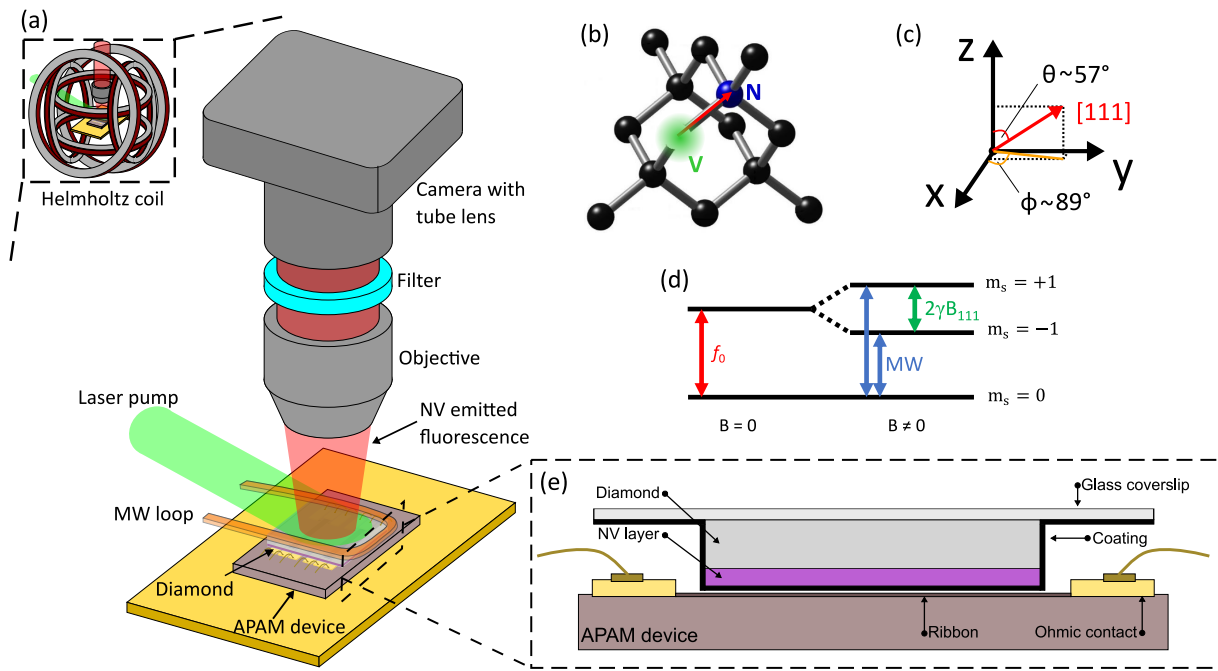}
\captionsetup{width=1\linewidth}
\caption{(a) Schematics of the NV wide-field magnetic microscope. (b) NV crystal structure, with the [111] direction indicated by the red arrow. (c) NV center [111] direction orientation in the reference frame set by the three Helmholtz coil axis. (d) NV center ground-state spin sublevels energy diagram, showing the zero-field splitting $f_0 \simeq 2.87$ GHz and the Zeeman effect lifting the $m_s = \pm$ 1 degeneracy. (e) Diamond-device integration method.}
\label{Figure1}
\end{figure*}

In this work, we used an ensemble of nitrogen-vacancy (NV) centers in diamond to map the magnetic fields generated by currents injected in an APAM device. For a proof-of-concept demonstration of NV magnetometry as an APAM materials diagnostic tool, we investigated a device made of straightforward micron-scale ribbon-shaped structures defined by mesas etched from blanket P-delta-doped Si. These structures are recognized, and utilized heavily, throughout the APAM community as a valid witness material for APAM doping processes~\cite{oberbeck2002encapsulation,hagmann2018high}. NV magnetometry~\cite{Doherty, Rondin, Chipaux, MicroMagnets} non-invasively measures stray magnetic fields by exploiting the magnetically-sensitive NV electronic spin ground-state, the ability to manipulate the sublevel populations using a resonant microwave field, and the optical spin readout by spin-dependent photoluminescence (PL). By collecting the NV fluorescence on a CMOS camera in a home-built optical microscope, we obtained a 2D magnetic field map over a few-millimeter field of view with micrometer-scale spatial resolution~\cite{QDM1ggg, edlynQDMreview}. We then reconstructed the current paths within the device from the measured magnetic images, revealing current flow features of a functioning APAM device, as well as that of a faulty one due to micrometer- and nanometer-scale material damage. This approach extends the suite of existing magnetic imaging tools used in the electronics failure analysis community (e.g.~scanning gate microscopy~\cite{ScanningGate}, scanning SQUID microscopy~\cite{felt_SQUID_MR}, and magnetic force microscopy~\cite{MFMshorts2}), with the added benefits of high spatial resolution, high instrument reliability, wide-field data collection, and non-invasive ambient-conditions operation.

\begin{figure*}[ht]
\centering
\includegraphics[width=.7\linewidth]{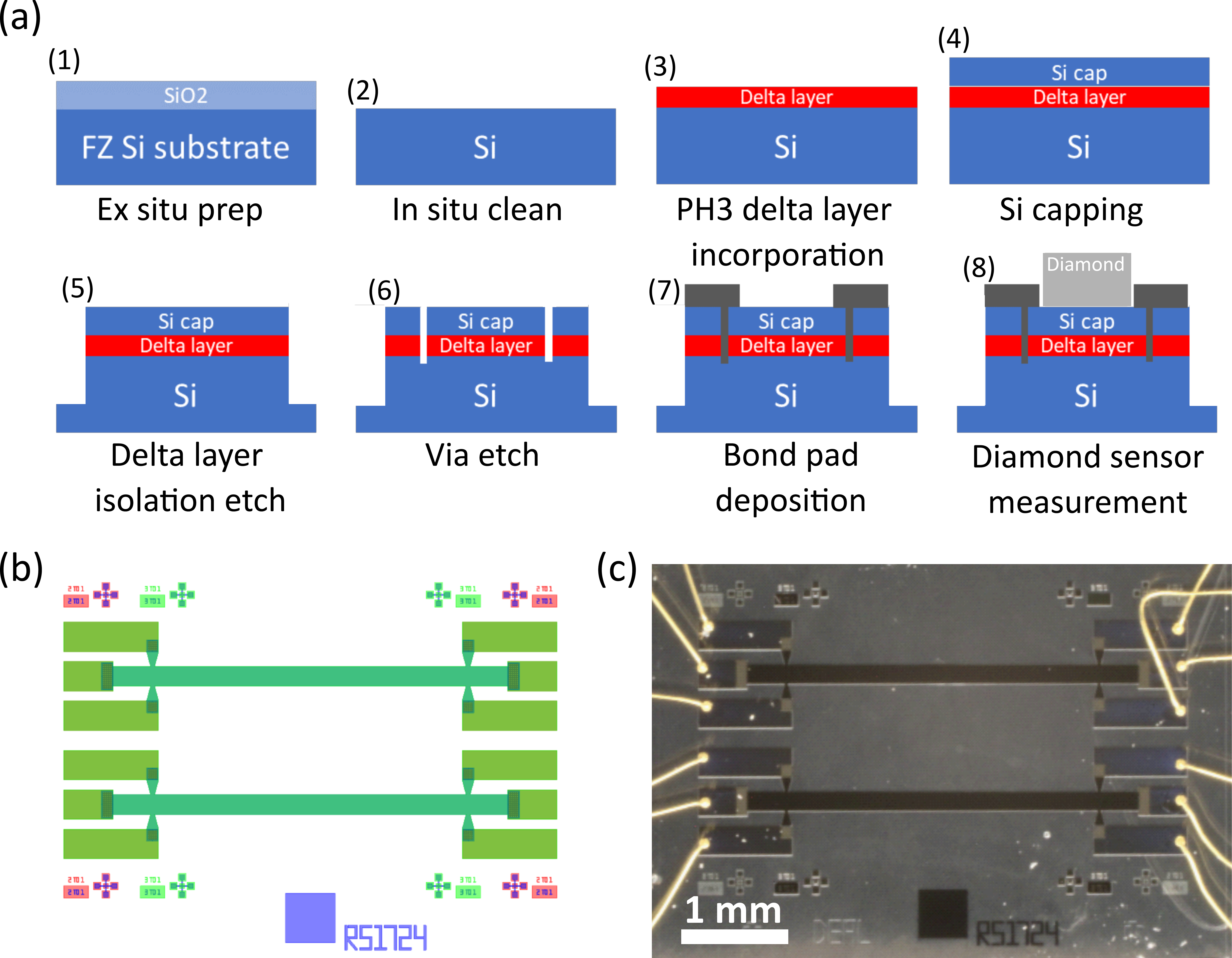}
\captionsetup{width=1\linewidth}
\caption{(a) APAM device fabrication process flow, shown in cross section. (b) Schematic view of the top side of the chip, showing the etched delta-doped layer having the shape of two parallel 200 $\upmu$m wide ribbons (dark green) and contact metallization (light green). (c) Final APAM device optical image.}
\label{Figure2}
\end{figure*}

\section{Materials and Methods}
\subsection{Experimental Setup}
To perform wide-field magnetic imaging, we used a home-built fluorescence microscope (schematic shown in figure~\ref{Figure1}(a))~\cite{edlynQDMreview}. The diamond sensor, placed on top of the APAM device, was illuminated from the side with a 532 nm laser beam, with a laser power of 1.1 W. The laser beam was expanded with a 400 mm focal length lens in order to obtain a uniform illumination in the camera field of view ($\sim$2.4$\times$3.8 mm$^2$). The NV fluorescence was collected by a 5x objective (0.13 numerical aperture) and filtered by a long-pass filter (650 nm cutoff wavelength) before being imaged trough a tube lens (f = 100 mm) onto a CMOS camera. The camera exposure time was set to 25 ms and a 4$\times$4 pixel binning was applied to obtain an image of 250$\times$400 pixels with a pixel size of 9.5 $\upmu$m. The microscope was placed inside a 3-axis Helmholtz coil set, used to apply a static $B_{0} \approx$ 1 mT bias magnetic field along the NV axis, defined as the [111] diamond crystallographic direction as shown in figure~\ref{Figure1}(b) and (c). The bias magnetic field lifts the degeneracy between the NV ground-state spin sublevels $m_S = \pm 1$ required to perform optically-detected magnetic resonance (ODMR) spectroscopy. The resonance frequencies of the [111] oriented NVs, as shown in figure~\ref{Figure1}(d), are shifted from the zero-field splitting $f_0 \simeq 2.87$ GHz by $\pm \gamma B_{111}$, with $\gamma \simeq 28$ kHz $\upmu$T$^{-1}$ being the NV gyromagnetic ratio and $B_{111}$ the magnetic field along the NV axis. As the laser pumps NVs into the $m_S=0$ bright sublevel, we performed ODMR spectroscopy by applying a microwave (MW) field to drive transitions from the $m_S = 0$ to the $m_S = \pm 1$ dark sublevels while monitoring the NV PL emission: when the MW is on resonance with the $m_S = \pm 1$, the PL intensity is reduced. In our apparatus, a copper loop placed above the diamond provided the MW excitation. The MW signal was generated by a TPI-1002-A MW source, pulsed by a Mini-Circuits ZASWA-2-50DRA+, then amplified into the copper loop with a Mini-Circuits ZHL-16W-43-S+. 

\subsection{NV Diamond Sample}
Our NV diamond sensor was a 4$\times$4$\times$0.5 mm$^3$ electronic grade diamond (native nitrogen density $<$ 5 ppb), that was CVD overgrown with a 4 $\upmu$m-thick $^{12}$C-enriched diamond layer doped with 25 ppm of $^{14}$N on one surface (process performed by Applied Diamond, Inc). After overgrowth, it was irradiated with a 1 MeV energy, 1.2e18 cm$^{-2}$ fluence electron beam and vacuum-annealed as in Ref.~\cite{Jake2022} to form a NV ensemble. Finally, the diamond was cleaned with triacid solution (1:1:1 sulfuric, nitric and perchloric acids) at 250 $^\circ$C for 1 hour. To avoid photocarriers excitation in the imaged device, we glued the non-NV-containing diamond side to a glass coverslip with a UV-curing transparent glue. Both the diamond and the coverslip were then coated with a three-layer stack made of 5 nm of Ti (to improve adhesion), 150 nm of Ag (to reflect photons and prevent photocurrent excitation), and 150 nm of Al$_2$O$_3$ (to avoid contact shorting in the device)~\cite{Pauli555}. As shown in figure~\ref{Figure1}(e), our measurements were performed with the diamond placed on top of the device, with the NV surface facing down. As a demonstration of the importance of protecting the device under studying, results of magnetic imaging without the three-layer stack and the consequences on APAM device functioning are reported in the supplementary materials~\cite{supplementary}. 

\subsection{Characterization Techniques}
SEM analysis was performed with a FEI Nanolab 650 microscope, operated in secondary electron mode with an acceleration voltage of $10$ keV and at a working distance of $5$ mm. Laser profilometry was done by using a Keyence VK-X150 microscope equipped with a 658 nm laser.

\begin{figure*}[ht]
\centering
\includegraphics[width=0.9\linewidth]{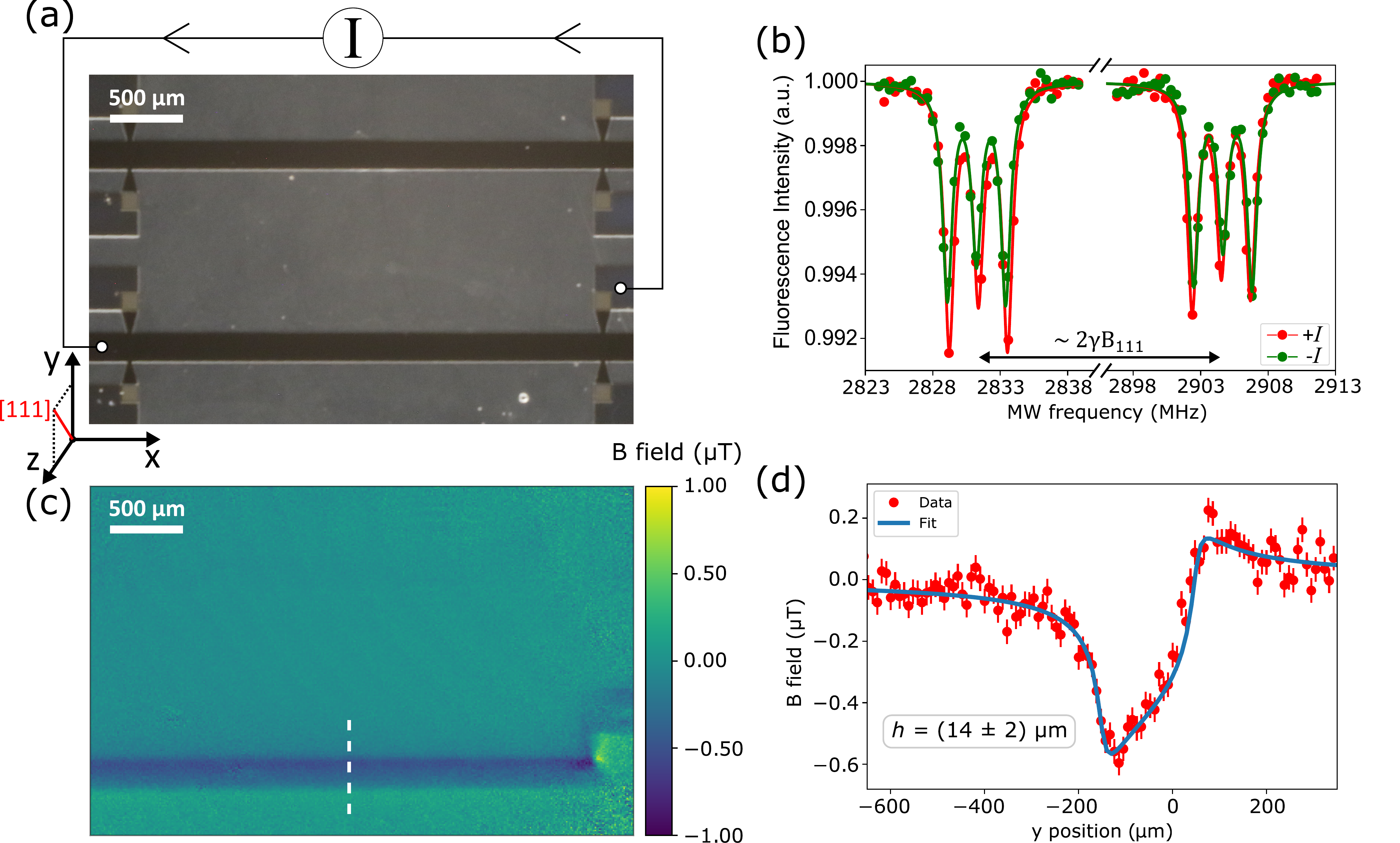}
\caption{(a) Optical image of the APAM device showing over which ribbon terminals a 150 $\upmu$A current was injected. (b) Typical ODMR spectrum for a single pixel for a positive (red dots) and negative (green dots) injected current. Each resonant transition is split into three peaks due to the $^{14}$N hyperfine interaction. The spectra are fit with six Lorentzian lineshapes (solid lines), yielding the value of $B_{111}$ for each  pixel. (c) NV magnetic field map for the same area shown in (a) obtained by $(B(+I)-B(-I))/2$. (d) Magnetic field profile along the white dashed line in (c) fit with Eq.~\ref{Bfit} (solid line) resulting in a stand-off distance of $h =(14$ $\pm$ $2)$ $\upmu$m.}
\label{Figure3}
\end{figure*}

\subsection{APAM Device}
The schematic of the APAM device fabrication process is shown in figure~\ref{Figure2}(a). APAM phosphorus delta-doped layer devices were prepared on flash-cleaned float zone Si(100) substrates through UHV dosing with phosphine and subsequent encapsulation with 30 nm of low temperature epitaxial silicon. The samples were then processed using standard cleanroom procedures using an all-optical lithography approach. Two Hall bar structures were etched into the silicon substrate through a Bosch etch, taking care to etch deep enough (10 $\upmu$m) to isolate the delta layer from the surrounding silicon substrate. An ICP silicon etch was then used to etch several shallow vias through the APAM silicon cap to the APAM delta layer. Finally, electrical contact to the APAM delta layer was achieved by the addition of 300nm-thick Ti/Al bond pads using a standard metals liftoff process. More details on the APAM device fabrication process are reported in Ref.~\cite{ward2017}. Figure~\ref{Figure2}(b) shows the schematic top view of the chip, whereas the final APAM device optical picture is shown in figure~\ref{Figure2}(c).

\section{Results and Discussion}
\subsection{Magnetic Field Measurements}
We injected a current of $I = 150$ $\upmu$A (applied voltage of 1.9 V) through one of the APAM device ribbons, as shown in figure~\ref{Figure3}(a). We performed wide-field ODMR spectroscopy, collecting emitted NV PL while sweeping the MW frequency, obtaining an ODMR spectrum for each camera pixel. We then fit the ODMR spectra for each pixel, shown in figure~\ref{Figure3}(b), to determine the resonance frequencies. The frequency difference between the two resonance frequencies $\Delta f$ is used to calculate the value of the magnetic field along the NV axis $B_{111} = \Delta f/2\gamma$ for each pixel, allowing us to obtain a 2D magnetic field map. To isolate the contributions of the injected current to the magnetic field (and remove any background magnetic fields), we collected two wide-field magnetic image measurements, one with positive current $I$, the other with a current of opposite sign $-I$. To improve the signal-to-noise ratio, each measurement was repeated and averaged 10 times for a total acquisition time of $\sim$3 h. The normalized difference of the two images $B_{111}(I) = (B(+I)-B(-I))/2$ was then calculated to remove the background magnetic field. The $B_{111}(I)$ magnetic field map resulting from wide-field ODMR spectroscopy is displayed in figure~\ref{Figure3}(c). 

\begin{figure*}[t]
\centering
\includegraphics[width=0.9\linewidth]{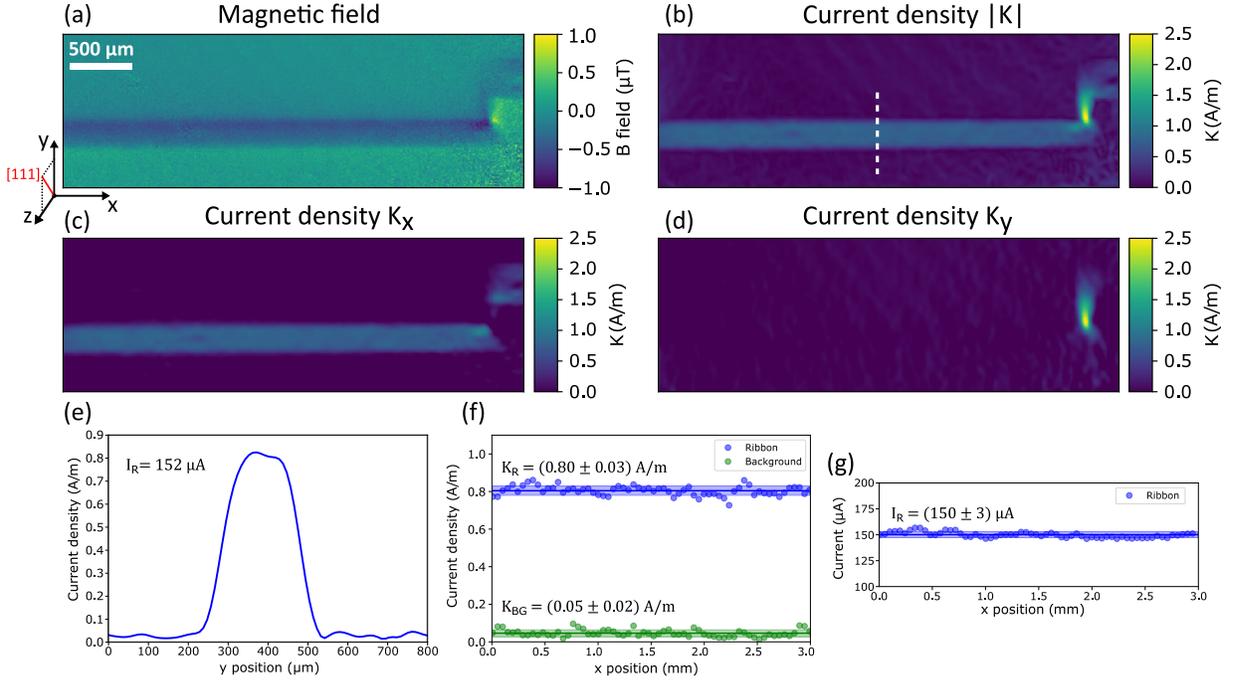}
\caption{(a) Magnetic field map for an injected current of $150~\upmu$A. (b) Reconstructed current density magnitude $|\vec{K}| = (K_x^2+K_y^2)^{-1/2}$. (c)-(d) $K_x$ and $K_y$ component respectively. The scale bar is the same for (a-d). (e) Current density along the white dashed line in (b), leading to an integrated current flowing in the ribbon of $I_\textrm{R} = 152$ $\upmu$A. (f) Surface current density as function of $x$ measured on a 1D line-cut in the center of the ribbon (blue dots) and in the background (green dots). The $K_R$ and $K_{BG}$ values are given by the average (solid lines) while the uncertainty is the standard deviation (shaded areas). (g) Current as function of $x$ inside the ribbon (blue dots) obtained by integrating $|\vec{K}|$ over the width of the ribbon. The $I_R$ values is given by the average (solid line) while the uncertainty is the standard deviation (shaded area).}
\label{Figure4}
\end{figure*}

\subsection{Stand-off Distance Estimation}
To estimate the stand-off distance $h$ between the NV-layer and the APAM device, we used the following equations for the magnetic field generated along a line-cut of an infinitely long 2D ribbon having width $w = 200$ $\upmu$m and carrying a current $I$~\cite{magFilm}:
\begin{equation}
\begin{split}
   & B_{y} =  \frac{\mu_0 I}{2 \uppi w} \left[\arctan\left(\frac{y-w/2}{h}\right) - \arctan\left(\frac{y+w/2}{h}\right)\right] , \\
    & B_{z} = \frac{\mu_0 I}{4 \uppi w} \log\left(\frac{h^2 + (y-w/2)^2}{h^2 + (y+w/2)^2}\right) .
\end{split}
\end{equation}
Here $y$ and $z$ are respectively the directions along the width of the ribbons and perpendicular to the device plane and $\mu_0 = 4\pi \times 10^{-7}$ m$\cdot$T/A. The magnetic field projection along the NV axis is given by:
\begin{equation}
B_{111} = B_{y}\sin\upphi\sin\uptheta+B_{z}\cos\uptheta ,
\end{equation}
with $\upphi \simeq$ 89$^{\circ}$ and $\uptheta\simeq$ 57$^{\circ}$ being the two angles defining the NV-axis direction. $\uptheta$ and $\upphi$ are respectively the angle from $z$ and the angle in the $xy$ plane, defining the direction of the bias magnetic field aligned with the [111] diamond axis, as shown in figure~\ref{Figure1}(c). 

Finally, the finite NV layer thickness $t_{NV} = 4$ $\upmu$m is taken into account by replacing $h$ with $z$, and averaging over the interval $h \leq z \leq h + t_{NV}$~\cite{Abrahams2021}:
\begin{equation}\label{Bfit}
    B_{NV} = \frac{1}{t_{\textrm{NV}}}\int_h^{h+t_{NV}}  B_{111}(y,z) ~ dz ,
\end{equation}
resulting in $h =(14$ $\pm$ $2)$ $\upmu$m, as shown in figure~\ref{Figure3}(d). The error bars on the experimental data in figure~\ref{Figure3}(d) come from the $\delta B$ standard deviation (noise floor) of the magnetic fields measured over a line-cut away from the ribbon, i.e.~where $\langle B_{111} \rangle = 0$, resulting in $\delta B = 0.04$ $\upmu$T. 

\subsection{Current Reconstruction}
After finding the stand-off distance, we reconstructed the surface current density $K(x,y)$ from the single vector component magnetic field map $B_{111}$. We assumed that the currents flowing in the APAM device are quasistatic and confined in a 2D sheet, and we used the inverted Biot-Savart law in Fourier space to obtain~\cite{wickswo89, degenCurrentImg, hollenbergCurrentImg}:
\begin{equation} \label{FT}
\begin{split}
  &\hat{K}_{x}(k_{x},k_{y}) = \frac{w(k, \lambda) k_{y}}{g(h, k_{z})[e_yk_y-e_xk_x+ie_zk]} b(k_{x},k_{y}, h), \\
  & \hat{K}_{y}(k_{x},k_{y}) = \frac{w(k, \lambda) k_{x}}{g(h, k_{z})[e_xk_x-e_yk_y-ie_zk]} b(k_{x},k_{y}, h).
\end{split}
\end{equation}
Here $\{\hat{K}_{x},\hat{K}_{y}\}$ and $b(k_{x},k_{y}, h)$ are respectively the current density components and the magnetic field map in Fourier space, $k_x$ and $k_y$ are the Fourier-space wave vector components, $k =(k_x^2+k_y^2)^{1/2}$ and $\{e_x,e_y,e_z\} = \{\sin\upphi\sin\uptheta, \cos\upphi\sin\uptheta, \cos\uptheta\}$ is the NV axis orientation. 
The function $g(k,z)$ is the Green's function
\begin{equation}
g(k,z) = \frac{\mu_0 }{2}e^{-kz},
\end{equation}
which requires a window function $w(k,\lambda)$ to suppress the measurement white noise at high spatial frequencies $k$ when $g(k,z)$ tends to zero. To achieve this, we use a Hann window function
\begin{equation}
w(k, k_{\textrm{max}} ) = 
\begin{cases}
\frac{1}{2}\left[ 1+ \cos \left(\pi k/k_{\textrm{max}} \right) \right] &  \text{if} \hspace{2mm} |k| < k_{\textrm{max}}\\
0 & \text{otherwise}
    \end{cases}
\end{equation}
with cutoff wave vector $k_{\textrm{max}}$. Noise rejection of high $k$ values reduces the spatial resolution of the calculated current distribution (which is already limited by the stand-off distance), so $k_{\textrm{max}}$ is found empirically to adjust the trade-off between these two effects. Finally, the inverse Fourier transform of Eq.~\ref{FT} leads to the current density components $K_x$ and $K_y$ as well as the current density magnitude $|\vec{K}| = (K_x^2+K_y^2)^{1/2}$.

\begin{figure*}[ht]
\centering
\includegraphics[width=0.9\linewidth]{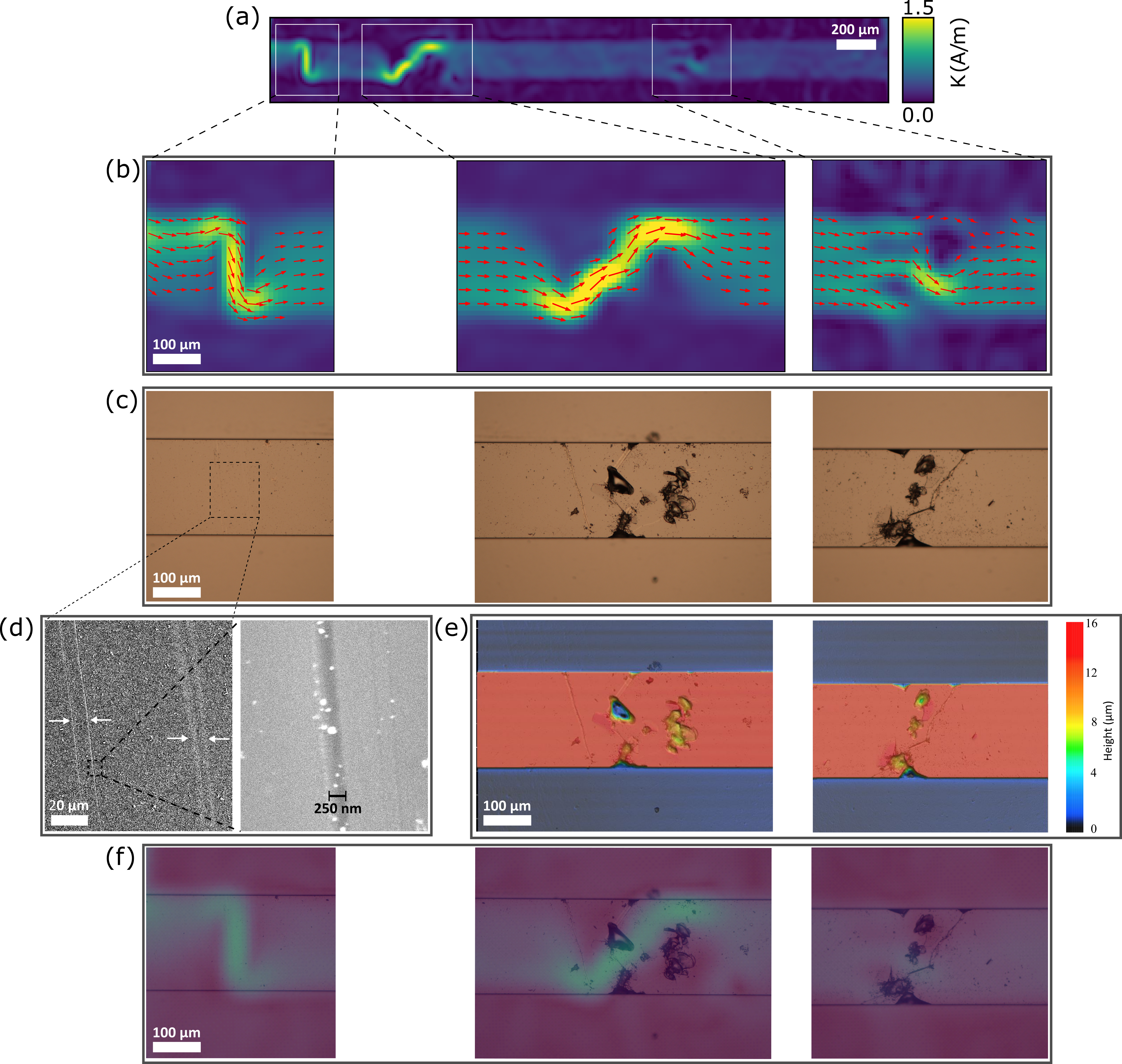}
\caption{(a) Reconstructed current density magnitude $|\vec{K}|$ of a damaged ribbon for an injected current of 150 $\upmu$A. (b) Zoom of areas of current flow irregularities. The arrows point in the direction of $\vec{K}$ with length proportional to $|\vec{K}|$. (c) Optical images corresponding to the same regions in (b). (d) SEM images with different magnification of the area in (c)-left, showing four scratches in the APAM device surface highlighted by the black arrows and a zoom of one of the scratches having a lateral size of $\simeq$ 250 nm. (e) Laser profilometry analysis of the same regions in the center and right images of (c). (f) Overlay of optical and $|\vec{K}|$ images. The scale bars for (b), (c), (e) and (f) are the same.}
\label{Figure5}
\end{figure*}

\subsection{Functioning APAM Device}
The reconstructed current density for the magnetic field map of figure~\ref{Figure4}(a) is shown in figure~\ref{Figure4}(b,c,d). To check the quality of the current reconstruction algorithm, we integrated the current density over the width of the ribbon to obtain a total current of $I_\textrm{R} = 152$ $\upmu$A (figure~\ref{Figure4}(e)). The value of the surface current density in the ribbon $K_R$ as function of $x$ is shown in figure \ref{Figure4}(f), resulting in a constant value of $K_R = (0.80 \pm 0.03)$ A/m. The resulting current obtained by integrating $K_R$ over the width of the ribbon is shown in figure \ref{Figure4}(g). The constant value of $I_R = (150 \pm 3)$ $\upmu$A proves the current stability across the P-doped region and matches the nominal injected current of 150 $\upmu$A. The background surface current density $K_{BG}$ measured in a region away from the ribbon is also shown in figure \ref{Figure4}(f), showing a mean offset baseline of $K_{BG} = 0.05$ A/m and a standard deviation of $\delta K = 0.03$ A/m. The DC offset $K_{BG}$ in the $K$ map is an artifact of the current reconstruction algorithm arising when $K$ is calculated from a single axis magnetic field measurement~\cite{hollenbergCurrentImg} while the standard deviation $\delta  = 0.03$ A/m represents the noise floor. $\delta K$ gives a measurement of the surface current density sensitivity, leading to a smallest detectable current (i.e.~current measured with a SNR = 1) in the 200 $\upmu$m wide ribbon of $\sim$6 $\upmu$A.

\begin{figure*}[ht]
\centering
\includegraphics[width=0.8\linewidth]{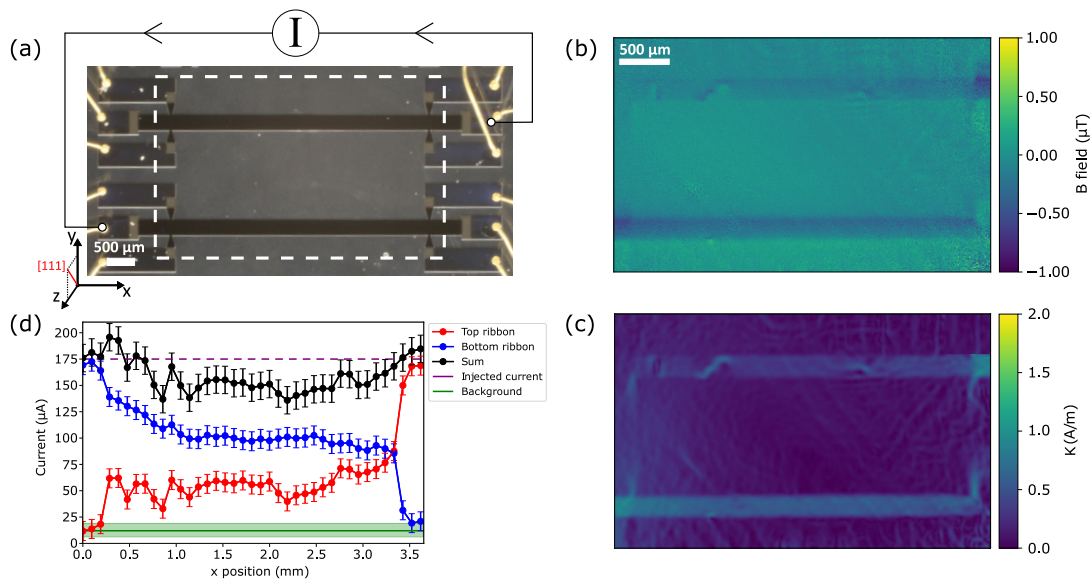}
\caption{(a) $I =$ 175 $\upmu$A current injection configuration. Dashed white box is the region imaged in (b) and (c). (b) Measured magnetic field map. (c) Reconstructed current density $|\vec{K}|$. The scale bar for (b) and (c) is the same. (d) Integrated current density for the top (red curve) and bottom ribbon (blue curve), with the sum between the two (black curve). The nominal injected current (purple curve) and the background (average value solid green line, standard deviation shaded green area) are also reported.}
\label{Figure6}
\end{figure*}

\subsection{Faulty APAM Device}
We can also detect failures in our APAM devices. An example is shown in figure~\ref{Figure5}(a) for an injected current of 150 $\upmu$A (applied voltage of $\sim$18 V), where there are sections of the ribbon for which we can observe impeded current flow. Figure \ref{Figure5}(b) shows a zoom of the current path irregularities: the current direction around holes and choke points are indicated by the arrows, which have lengths proportional to the local current density magnitude. To asses the origins of these irregularities in the current flow, we compared the measured current density maps with optical images of the same areas (figure~\ref{Figure5}(c)), as well as SEM (figure~\ref{Figure5}(d)) and laser profilometry (figure~\ref{Figure5}(e)) analysis. From the set of optical images in figure~\ref{Figure5}(c) we could observe a correlation between current flow irregularities and material damage on the APAM device surface. SEM analysis on the left image of figure~\ref{Figure5}(d) shows the presence of two pairs of scratches each with lateral size of $\simeq$ 250 nm, as shown in right panel of figure~\ref{Figure5}(d). We then used laser profilometry to obtain a 3D altitude image of the same areas. As seen in figure~\ref{Figure5}(e) some of the cracks on the device surface are several microns deep, enough to have removed the conductive P-doped region below Si surface. This allows us to attribute the current flow irregularities to the presence of material damage, such as cracks and scrapes. The overlay between current density maps and optical images, shown in figure~\ref{Figure5}(f), helps to see where the current is flowing around the device defects. 

\subsection{Current Leakage Between APAM structures}
Finally, with the aim of detecting current leakage paths, we intentionally created leakages between the P-doped ribbons by connecting both of them to the current source, as shown in figure~\ref{Figure6}(a). By doing this, we expect the current to leave the bottom ribbon and flow through the Si substrate to reach the top one. Unsurprisingly, to inject a current of $I = 175$ $\upmu$A we need to apply a voltage of $\sim$70 V, $\sim$10x higher than the previous cases, consistent with the current flowing through an undoped path. The resulting magnetic field map and reconstructed current density, reported in figure~\ref{Figure6}(b) and (c) respectively, show that the current is flowing in both ribbons. For a quantitative analysis we integrated the current density over the ribbons to calculate the current flowing along $x$, with the result shown in figure~\ref{Figure6}(d). The error bars on the integrated current are calculated starting from the uncertainty on the measured magnetic field $\delta B = 0.04$ $\upmu$T, that is propagated through the current reconstruction algorithm to obtain $\delta I = 9$ $\upmu$A. We observed that the current profiles are consistent with the current being injected in the bottom ribbon and collected from the top one. Interestingly, the sum of the currents flowing in the two ribbons matches the total injected current only close to the injection and collection points, implying that there are leakage currents in between the ribbons that we do not detect. More quantitatively, the unobservable current, i.e.~leakage current defined as the difference between the nominal injected current and total measured current (black curve in figure~\ref{Figure6}(d)), is at most 39 $\pm$ 13 $\upmu$A. As the measured surface current density sensitivity is $\delta K = 0.03$ A/m (see figure~\ref{Figure4}(f)), the leak current to be detected with a SNR = 1 must flow in a sheet less than $\sim$1 mm wide. We postulate that we could not directly observe the current leakage path for several reasons. The first explanation is that the leakage path is three dimensional, making the current density reconstruction unreliable due to the 2D current density assumption of our model. The second explanation is that the leakage currents, no longer confined in the 2D P-doped regions, are spread over a wide area, resulting in a surface current density smaller than $\delta K$. Finally, if the leakage current behaves as a uniform sheet between the two ribbons, then its field would be largely along the $x$ direction (except at the edges), which is orthogonal to the NV sensing direction and would be largely invisible. Although the field from such a leakage current would be too weak to detect with the present apparatus, a follow-up experiment with an improved $\delta K$ that measures the vector magnetic fields for each pixel may be able to detect it.

\section{Conclusion}
In summary, we have shown the capability to map the magnetic fields in an APAM device with micron-scale resolution over a millimeter-scale field of view area using NV wide-field magnetic imaging under ambient conditions without interfering with the operation of the device under study (see supplementary materials~\cite{supplementary}). From the measured magnetic field maps, we reconstructed the surface current densities, leading to wide-field imaging of the current flowing in the device. That allowed us to obtain information on the APAM device properties and detect unexpected behavior, such as current path irregularities and leakages. Current density sensitivity is of order $0.03$ A/m, leading to a minimum detectable current of $\sim$6 $\upmu$A in the 200 $\upmu$m wide APAM ribbon. Further work will upgrade this experiment to detect and reconstruct 3D currents~\cite{Levine3D} as well as AC signals in the MHz or GHz range with appropriate NV manipulation techniques~\cite{maletinskyMWimger, dima_magReview}. The development of a NV-based diagnostic tool able to image where current is flowing and detect failures or leakages down to $0.03$ A/m could benefit the whole microelectronics community, with application not limited to APAM devices but extended to CMOS based technologies~\cite{qdmFPGA}. 

\section*{Acknowledgements}
We thank George Burns, Michael Titze, and Edward Bielejec (Ion Beam Laboratory, Sandia National Laboratories) for help with diamond preparation. Sandia National Laboratories is a multi-mission laboratory managed and operated by National Technology and Engineering Solutions of Sandia, LLC, a wholly owned subsidiary of Honeywell International, Inc., for the DOE's National Nuclear Security Administration under contract DE-NA0003525. This work was funded, in part, by the Laboratory Directed Research and Development Program and performed, in part, at the Center for Integrated Nanotechnologies, an Office of Science User Facility operated for the U.S.~Department of Energy (DOE) Office of Science. This paper describes objective technical results and analysis. Any subjective views or opinions that might be expressed in the paper do not necessarily represent the views of the U.S. Department of Energy or the United States Government. 

\section*{References}
\bibliography{Main}

\end{document}


\title{Supplemental material for Current Paths in an Atomic
Precision Advanced Manufactured Device Imaged by
Nitrogen-Vacancy Diamond Magnetic Microscopy}

\date{\today}
\author{Luca Basso}\affiliation{\cint}
\author{Pauli Kehayias}\affiliation{\snl}
\author{Jacob Henshaw}\affiliation{\cint}
\author{Maziar Saleh Ziabari}\affiliation{\cint}\affiliation{\unm}
\author{Heejun Byeon}\affiliation{\cint}
\author{Michael P. Lilly}\affiliation{\cint}
\author{Ezra Bussmann}\affiliation{\snl}
\author{Deanna M. Campbell}\affiliation{\snl}
\author{Shashank Misra}\affiliation{\snl}
\author{Andrew M. Mounce}\affiliation{\cint}

\maketitle

\subsection*{Diamond sensor reflective coating}
\begin{figure}[ht]
\centering
\includegraphics[width=1\linewidth]{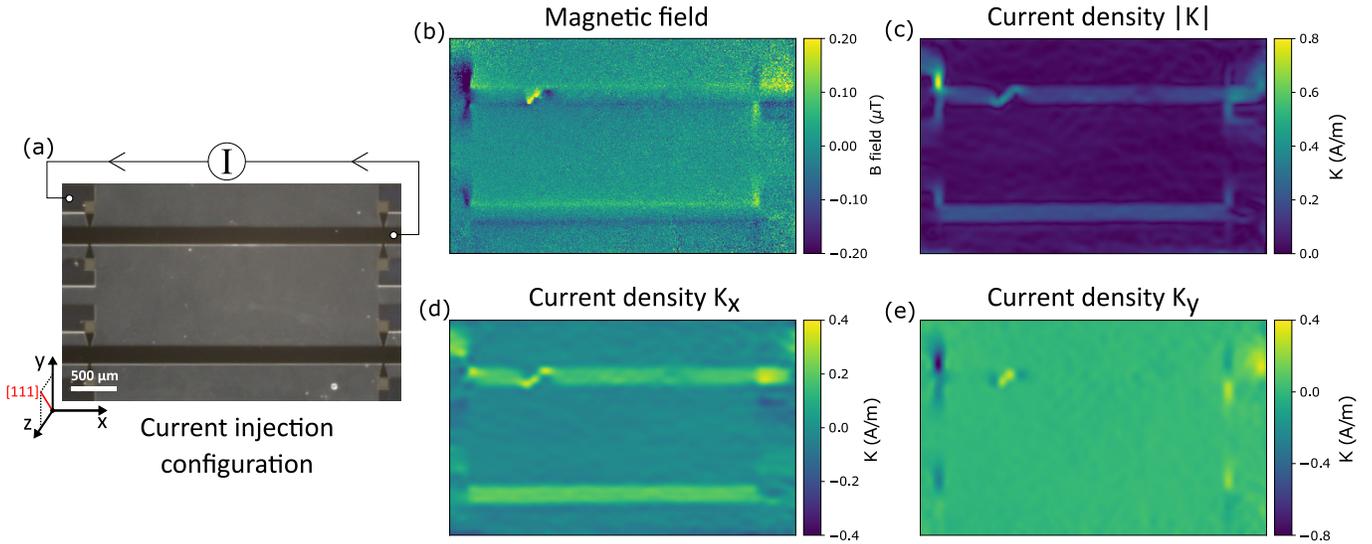}
\captionsetup{width=1\linewidth}
\caption{Magnetic field measurement without diamond protective coating. (a) APAM device optical picture showing over which ribbon terminals a 100 $\upmu$A current is injected. (b) Measured magnetic field map. (c) Reconstructed surface current density magnitude $|\vec{K}| = (K_x^2+K_y^2)^{1/2}$. (d)-(e) $K_x$ and $K_y$ components respectively. The scale bar is the same for (a)-(e).}
\label{FigureS1}
\end{figure}
To avoid creating photoexcited carriers in the APAM device from the NV pump laser impinging on the Si substrate, we deposited on the NV-center enriched diamond surface a three-layer stack reflective coating (see section 2.2 in the main text). To highlight the importance of the protective coating, we performed the same magnetic field measurement with the uncoated diamond. Figure~\ref{FigureS1} shows a typical magnetic field and reconstructed current maps obtained with the uncoated diamond when the current is injected only in the top ribbon, as shown in figure~\ref{FigureS1}(a). Despite injecting the current in only one ribbon, we could clearly also observe the current flowing in the unplugged bottom ribbon, both in the magnetic field (figure~\ref{FigureS1}(b)) and in the reconstructed surface current density maps (figure~\ref{FigureS1}(c)-(e)). This is attributed to laser going through the diamond and impinging on the below APAM device Si surface, leading to free-carrier excitation and a consequent presence of photocurrents in the device. For comparison, figure 3(c) in the main text does not show any current leaving the ribbon and flowing in the unplugged ribbon when the coated diamond is used for the experiment. The presence of photocurrents is also confirmed by simultaneously measuring the resistance during magnetic field imaging. Across the top ribbon, as in figure~\ref{FigureS1}(a), the resistance drops from R $\approx$ 120 k$\Omega$ to R $\approx$ 16 k$\Omega$ when the laser is turned on and the diamond is uncoated, whereas no relevant change in R is observed when the diamond is coated.